\documentstyle[twoside,fleqn,npb,epsfig,psfig]{article}
\newcommand{\be}{\begin{equation}}
\newcommand{\ee}{\end{equation}}
\newcommand{\bea}{\begin{eqnarray}}
\newcommand{\eea}{\end{eqnarray}}
\newcommand{\bean}{\begin{eqnarray*}}
\newcommand{\eean}{\end{eqnarray*}}
\newcommand{\ba}{\begin{array}}
\newcommand{\ea}{\end{array}}

\newcommand{\AmS}{{\protect\the\textfont2
  A\kern-.1667em\lower.5ex\hbox{M}\kern-.125emS}}

\title{Lepton Asymmetry in Polarized Drell-Yan
       \thanks{Presented by J.K. at RADCOR 2002 and Loops and Legs 2002, 
                Kloster Banz (Germany), September 8 to 13, 2002.}}

\author{Jiro Kodaira\address{
        Department of Physics, Hiroshima University,
        Higashi-Hiroshima, 739-8526, Japan} and Hiroshi Yokoya\address{
        Department of Physics, Hiroshima University,
        Higashi-Hiroshima, 739-8526, Japan}}%

\begin{document}

\begin{abstract}
The lepton helicity distributions
in the polarized Drell-Yan process at RHIC energy
are investigated.
In the absence of the weak interaction, only
the measurement of lepton helicity can prove the antisymmetric part
of the hadronic tensor.
Therefore it might be interesting to consider
the helicity distributions of leptons to obtain more information
on the structure of nucleon from the polarized Drell-Yan process.
We estimate the QCD corrections at ${\cal O} (\alpha_s)$ level to the
hadronic tensor including both intermediate $\gamma$ and $Z$ bosons.
We report the numerical analyses on the $Z$ pole
and show that the $u (\bar{u})$ and $d (\bar{d})$ quarks give different
and characteristic contributions to the lepton helicity distributions.
We also estimate the lepton helicity asymmetry
for the various proton's spin configurations.
\end{abstract}

\maketitle

\section{INTRODUCTION}

In the last ten years, great progress has been made
both theoretically and experimentally in hadron spin physics.
Furthermore, in conjunction with 
new projects like the \lq\lq RHIC spin project\rq\rq,
\lq\lq polarized HERA\rq\rq , etc.,  we are now
in a position to obtain more information on the 
spin structure of nucleons.
The spin dependent quantity is, in general, very sensitive to the
structure of interactions among various particles. 
Therefore, we will be able to study the detailed structure of hadrons
based on QCD.
We also hope that we can find some clue to new physics beyond the
standard model through the new experimental data.

It is now expected that the polarized proton-proton collisions
(RHIC-Spin) at BNL relativistic heavy-ion collider RHIC\cite{RHIC} 
will provide sufficient experimental data to unveil the 
structure of nucleon. 
Therefore it is important and interesting to investigate various 
processes which might be measured in RHIC
polarized proton-proton collisions.

In this talk, we report the QCD one-loop calculations of
lepton helicity distributions from the polarized Drell-Yan process.
The lepton helicity distributions carry more information
on the nucleon structure than the \lq\lq inclusive\rq\rq\ Drell-Yan
observable like the $Q^2$ (invariant mass of leptons) dependence
of the cross section.
We will show that the $u (\bar{u})$ and $d (\bar{d})$ quarks give
characteristic contributions to the lepton helicity distributions.

\section{LEPTON HELICITY DISTRIBUTION}

The polarized Drell-Yan process as well as $W,Z$ production
has been studied by many authors
both for longitudinally\cite{LDR}
and transversely\cite{TDR} polarized case.
Thanks to the factorization theorem, the the Drell-Yan cross section
is given as the convolution of parton densities $f_i (x)$ with
hard subprocess cross section $d \hat{\sigma}_{ij}$,
\[ d \sigma = \int dx_1 dx_2 \sum_{ij} \ d \hat{\sigma}_{ij}
               \ f_i (x_1) f_j (x_2) \ .\]
Now let us consider, for simplicity, the 
virtual $\gamma$ mediated Drell-Yan process.
The subprocess cross section $d \hat{\sigma}$ is
written in terms of the hadronic and leptonic tensors as,
\bean
   d \hat{\sigma} &\propto&
      \left( \hat{W}_{\mu\nu}^S + \hat{W}_{\mu\nu}^A \right)
          \left( L^{S\, \mu\nu} + L^{A\, \mu\nu} \right)\\
      &=& \hat{W}_{\mu\nu}^S L^{S\, \mu\nu}
           + \hat{W}_{\mu\nu}^A  L^{A\, \mu\nu} \ .
\eean
The anti-symmetric part of hadronic tensor $W_{\mu\nu}$
contains spin information on the annihilating partons.
However, for observables obtained after integrating out
the lepton distributions, this anti-symmetric part drops out.
Furthermore, the chiral structure of QED and QCD interactions
tells us that only particular helicity states are selected
for the $q-\bar{q}$ annihilation.
This observation shows that the polarized and unpolarized
Drell-Yan processes are governed by the essentially the same
dynamics at least for the hard part.

On the other hand, if we measure the lepton helicity
distributions,  we can reveal the whole structure of the
hadronic tensor as you can see {\it e.g.} in the tree level
result for the parton cross section,
\bean
  \lefteqn{\frac{d\hat{\sigma}^{(0)} (\lambda \,,\, \lambda_l)}
                    {dQ^2 d \cos \theta}}\\
  &=& \frac{\pi \alpha^2}{6Q^4} e^2_q
    \left(1 + \cos^2 \theta + 2 \lambda \lambda_l \cos \theta
     \right) \delta ( 1 - z )\ ,
\eean
where $z = Q^2 / \hat{s}$, $\theta$ is the scattering angle
of produced lepton in the partons CM frame
and $\lambda (\lambda_l )$ is the helicity of quark (lepton).
The third term comes from $\hat{W}_{\mu\nu}^A$.

\section{QCD ONE-LOOP CALCULATION}

At the QCD one-loop level, infrared and mass singularities appear
and we regularize them by giving a non-zero mass $\kappa$
to gluon\cite{MGR}.
To perform the numerical analyses by using {\it e.g.}
the $\overline{\rm MS}$
parameterization for the parton densities, 
we have to change the scheme.
However, it is well known\cite{scheme} how to do it.
We calculate the differential cross section for the lepton,
\[  \frac{d \hat{\sigma}}{dQ^2d cos\theta} \ ,\]
with helicities of partons and produced lepton being fixed.
In the following expressions, we suppress the complexity
coming from the two intermediate states of $\gamma$ and $Z$
boson.

The virtual gluon correction

\vspace{-0.8cm}
\begin{figure}[h]
\begin{center}
\begin{tabular}{cc}
\leavevmode\psfig{file=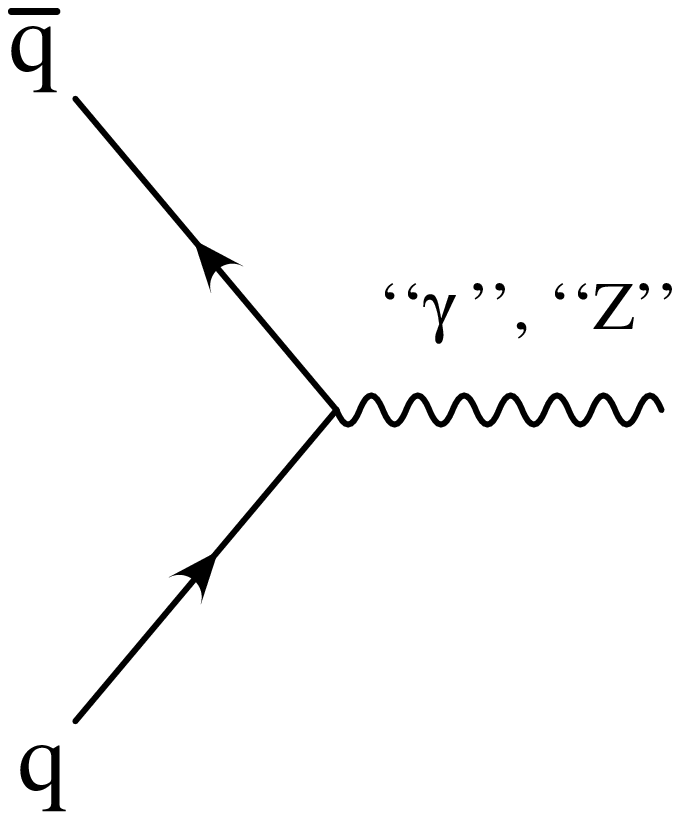,height=1.5cm} \qquad & \qquad
\leavevmode\psfig{file=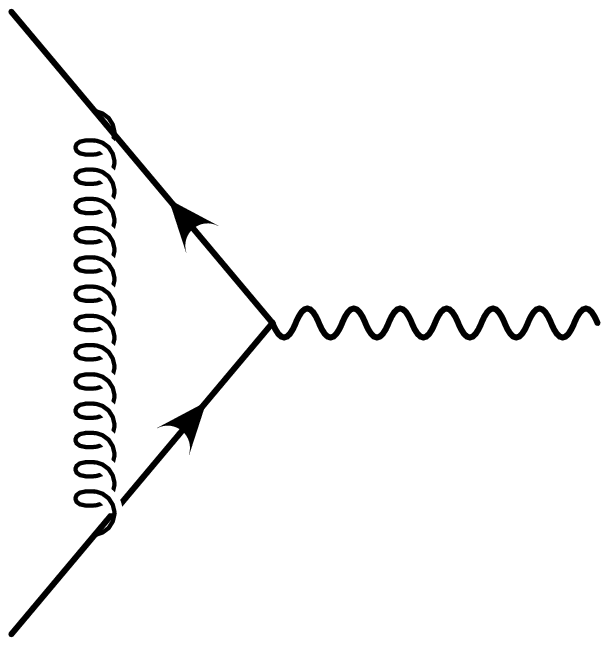,height=1.5cm}
\end{tabular}
\end{center}
\end{figure}

\vspace{-1.2cm}
\noindent
reads,
\bean
  \lefteqn{\frac{d\hat{\sigma}^{V}(\lambda , \lambda_l )}
          {dQ^2 d \cos \theta} = 
    \frac{d\hat{\sigma}^{(0)} (\lambda , \lambda_l )}
           {dQ^2 d\cos \theta} \ \delta (1 - z )}\\
  &\times& \left[ 1 + \left( \frac{\alpha_s}{\pi} C_F \right) \right. \\
  && \times \left.  \left( - \frac{1}{2}\ \ln^2 \frac{Q^2}{\kappa^2}
       + \frac{3}{2}\ \ln \frac{Q^2}{\kappa^2}
       - \frac{7}{4} + \frac{\pi^2}{6} \right) \right]\ .
\eean
The correction due to real gluon emission

\vspace{-0.8cm}
\begin{figure}[h]
\begin{center}
\begin{tabular}{cc}
\leavevmode\psfig{file=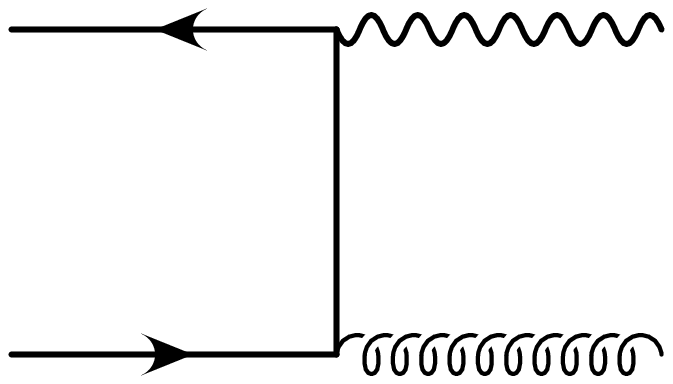,height=1.2cm} \qquad & \qquad
\leavevmode\psfig{file=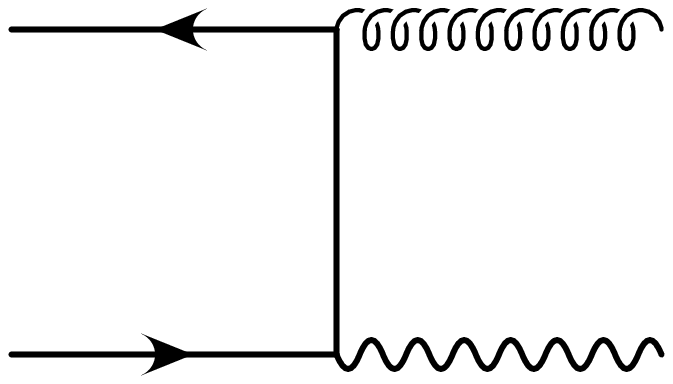,height=1.2cm}
\end{tabular}
\end{center}
\end{figure}

\vspace{-1cm}
\noindent
yields,
\bean
 \lefteqn{\frac{d\hat{\sigma}^{R} (\lambda , \lambda_l )}
       {dQ^2 d \cos \theta} = 
   \frac{d\hat{\sigma}^{(0)} (\lambda , \lambda_l )}
   {dQ^2 d\cos \theta} \ LB^R \ 
        \left(\frac{\alpha_s}{\pi} C_F \right)}\\
  &\times& \left[ \delta (1- z) 
     \left( \, \frac{1}{2} \, \ln^2 \frac{Q^2}{\kappa^2}
  - 2 \, \ln^2{2} \right) \right.\\
  &&  + \frac{1 + z^2}{(1 - z)_+}
     \ln \frac{Q^2}{\kappa^2} - 2 (1 + z^2) \frac{\ln{z}}{1 - z} \\
  && \left. + 2 (1 + z^2)
  \left( \frac{\ln (1 - z)}{1 - z} \right)_+ -(1-z) \right]\\
  &+&  \ DY^R (z, \cos\theta ; \lambda,\lambda_l) \ .
\eean
Finally, the contribution from the quark-gluon Compton
process

\vspace{-0.8cm}
\begin{figure}[h]
\begin{center}
\begin{tabular}{cc}
\leavevmode\psfig{file=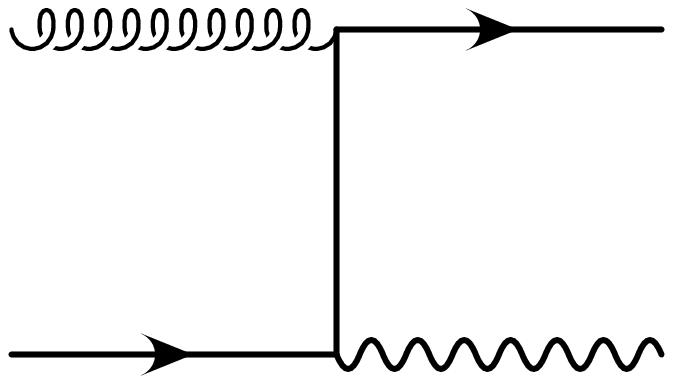,height=1.3cm} \qquad & \qquad
\leavevmode\psfig{file=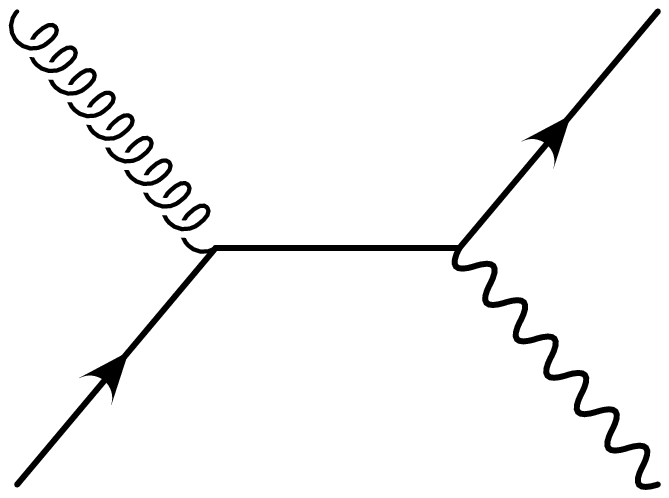,height=1.3cm}
\end{tabular}
\end{center}
\end{figure}

\vspace{-1cm}
\noindent
takes the form,
\bean
  \lefteqn{\frac{d\hat{\sigma}^{C} (\lambda, h, \lambda_l)}
       {dQ^2 d\cos \theta}
    = \frac{d\hat{\sigma}^{(0)} (\lambda , \lambda_l)}
         {dQ^2 d\cos \theta} \ LB^C \  
         \frac{\alpha_s}{2\pi} P_{qg} (z; \lambda , h)}\\
   &\times& \left[ \ln \frac{Q^2}{\kappa^2}
    + \ln \left (\frac{1-z}{z^2} \right) \right.\\
   && \qquad + \left. \ln 
     \left( \frac{(1+z- (1-z)\cos\theta)^2}{4z} \right) \right]\\
  &+&  \ DY^C (z, \cos\theta ; \lambda, h, \lambda_l) \ ,
\eean
where $h$ is the gluon helicity.

In the above two equations, $LB^{R,C}$ are the Lorentz boost
factors which are the function of $z,\cos\theta$ and helicities of
involved particles.
$DY^{R,C}$ are the finite contributions.
The explicit forms of these functions are very complicated
and lengthy\cite{KY} to be presented here.   
After combining all contributions, the double logarithmic singularities
cancel out and only mass singularities remain.
The coefficients of them are exactly
the DGLAP one-loop splitting functions $P_{qq}$ and $P_{qg}$. 

\section{NUMERICAL RESULTS}

By performing the factorization of mass singularities 
with an appropriate scheme transformation and convolution
integral of hard part with the parton distribution functions,
we can predict the helicity distributions of lepton
in proton-proton annihilation.
\bean 
 \lefteqn{\frac{d\sigma (\lambda_A, \lambda_B, \lambda_l)}
     {dQ^2 d\cos\theta^*}}\\
  &=&  \frac{1}{4}\sum_{f,\lambda} \int
      dx_A dx_B \left(q_f(x_A) + \lambda\lambda_A \Delta
      q_f(x_A) \right)\\
  && \times \quad \left(\bar{q}_f(x_B) - \lambda\lambda_B  \Delta
      \bar{q}_f(x_B) \right)\\
  && \times  \left[\frac{d\hat{\sigma}^{(0)}_f (\lambda,\lambda_l)}
       {dQ^2 d\cos\theta^*} +
    \frac{d\hat{\sigma}_f^{V+R,F} (\lambda, \lambda_l)}
     {dQ^2 d\cos\theta^*} \right]\\
  &+& \frac{1}{4}\sum_{f,\lambda,h}
   \int dx_A dx_B \left(q_f(x_A) + \lambda\lambda_A\Delta
      q_f(x_A) \right)\\
  && \times \left( g(x_B) +h \lambda_B \Delta
      g(x_B) \right)
     \frac{d\hat{\sigma}^{C,F}_f (\lambda, h, \lambda_l)}
      {dQ^2 d\cos\theta^*}\\
  &+& \ (A \leftrightarrow B) \, ,
\eean
where $\lambda_A$ and $\lambda_B$ are the helicities of
annihilating protons, $\cos\theta^*$ is the scattering
angle of the lepton in the protons CM frame
and the $Q^2$ dependence of parton densities is suppressed.
Note that $\hat{\sigma}^{R,F}$ and $\hat{\sigma}^{C,F}$
differ from $DY^R$ and $DY^C$ in the previous section
by the appropriate factor depending on the change of scheme.

Although our formulae can be applied for arbitrary total energy $s$
and invariant mass of lepton pairs $Q^2$,
we report in this talk only the results for $\sqrt{s} = 500$ GeV
and $Q = 91.2$ Gev (on $Z$-boson pole).
We use the $\overline{\rm MS}$ parameterization of
the parton densities in Ref.\cite{PDF1}.

Denoting the helicities of initial protons by $P_{A,B}(\pm)$,
there are three cases corresponding to $P_A (+) P_B (-)$,
$P_A (+) P_B (+)$, $P_A (-) P_B (-)$ configurations.
Among these, we plot in Fig.1 the most interesting
case in which we can discriminate the contributions
from $u(\bar{u})$ and $d(\bar{d})$ quarks.
Fig.1 shows the negative helicity lepton distributions in [pb/GeV]
with $P_A (+) P_B (-)$ configuration.
We have changed the lepton variable from the scattering angle to
the rapidity $y_l$.
The solid line is the total contribution.
The double line (solid and dashed) which has a peak
in the negative (positive) rapidity region is the contribution
from $u(\bar{u})$ ($d(\bar{d})$) quark.
The tree level predictions are displayed by the dot-dashed lines.

\vspace{-0.8cm}
\begin{figure}[h]
\begin{center}
\leavevmode\psfig{file=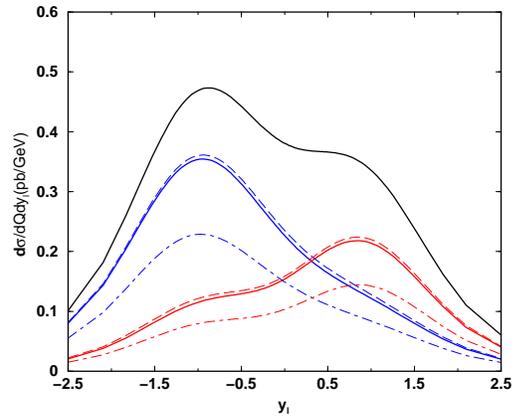,width=6.7cm}
\vspace{-1cm}
\caption{Negative helicity lepton distribution with $P_A (+) P_B (-)$}
\label{fig:1}
\end{center}
\end{figure}

\vspace{-1cm}
Some comments are in order for this result.
Firstly, the main effect of QCD correction is just an enhancement
of the tree level cross section: the $K$ factor is
$1.5 \sim 1.8$.
It does not change significantly the shape of the lepton
distributions.
Secondly, the different contributions from 
$u(\bar{u})$ and $d(\bar{d})$ quarks can be understood
intuitively by observing the following aspects:
(1) The polarized quark distributions in the polarized proton
$P(+)$ tell us,
\[ u (\uparrow) \gg d (\downarrow) \sim u (\downarrow)
           \gg d (\uparrow) \gg \bar{u} (\uparrow , \downarrow)
         \sim  \bar{d} (\uparrow , \downarrow) \ ,\]
where ${\uparrow}, {\downarrow}$ mean the quark's spin parallel and
anti-parallel to the parent proton's spin.
This relation implies the dominant subprocesses for the $P_A (+) P_B (-)$
case to be (i) $u_A (\uparrow) \bar{u}_B$, (ii) $\bar{u}_A u_B (\uparrow)$,
(iii) $d_A (\downarrow) \bar{d}_B$ and (iv) $\bar{d}_A d_B (\downarrow)$.
(2) From the angular momentum conservation,
the spin of produced $Z$ boson is aligned to $P_A$ ($P_B$) direction
for $u_A \bar{u}_B$ and $\bar{u}_A u_B$ 
($d_A \bar{d}_B$ and $\bar{d}_A d_B$) annihilations.
Furthermore, it is well known that the negative (positive)
helicity lepton from the $Z$ decay has higher probability to be produced
in the opposite (same) direction of $Z$ boson's spin.  
(3) The third point to be noted is that the $V-A$ coupling is larger
than the $V+A$ coupling for the quark and $Z$ boson interaction.
This suggests that among four subprocesses in (1),
(ii) and (iii) eventually dominate the process.
(4) Finally, since the momentum fractions of quarks are bigger than those of
anti-quarks, the distributions of negative helicity lepton
from (ii) and (iii) are Lorentz boosted to the negative 
and positive rapidity regions respectively.

\vspace{-1cm}
\begin{figure}[h]
\begin{center}
\leavevmode\psfig{file=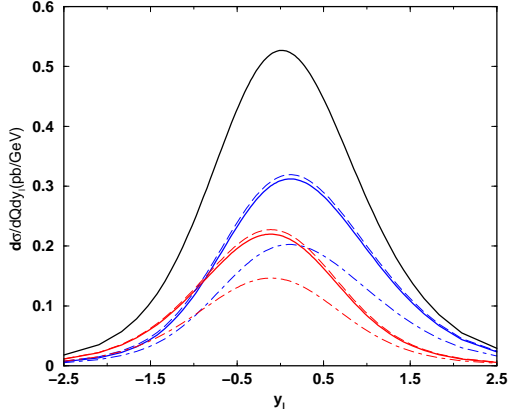,width=6.7cm}
\vspace{-1cm}
\caption{Positive helicity lepton distribution with $P_A (+) P_B (-)$}
\label{fig:2}
\end{center}
\end{figure}

\vspace{-1cm}
We plot in Fig.2 the positive helicity lepton distributions
with $P_A (+) P_B (-)$ configuration.
In this case, we do not see such a characteristic feature
like one in Fig.1.
The $u \bar{u}$ contribution slightly dominates the cross section.

\vspace{-0.8cm}
\begin{figure}[h]
\begin{center}
\begin{tabular}{cc}
\leavevmode\psfig{file=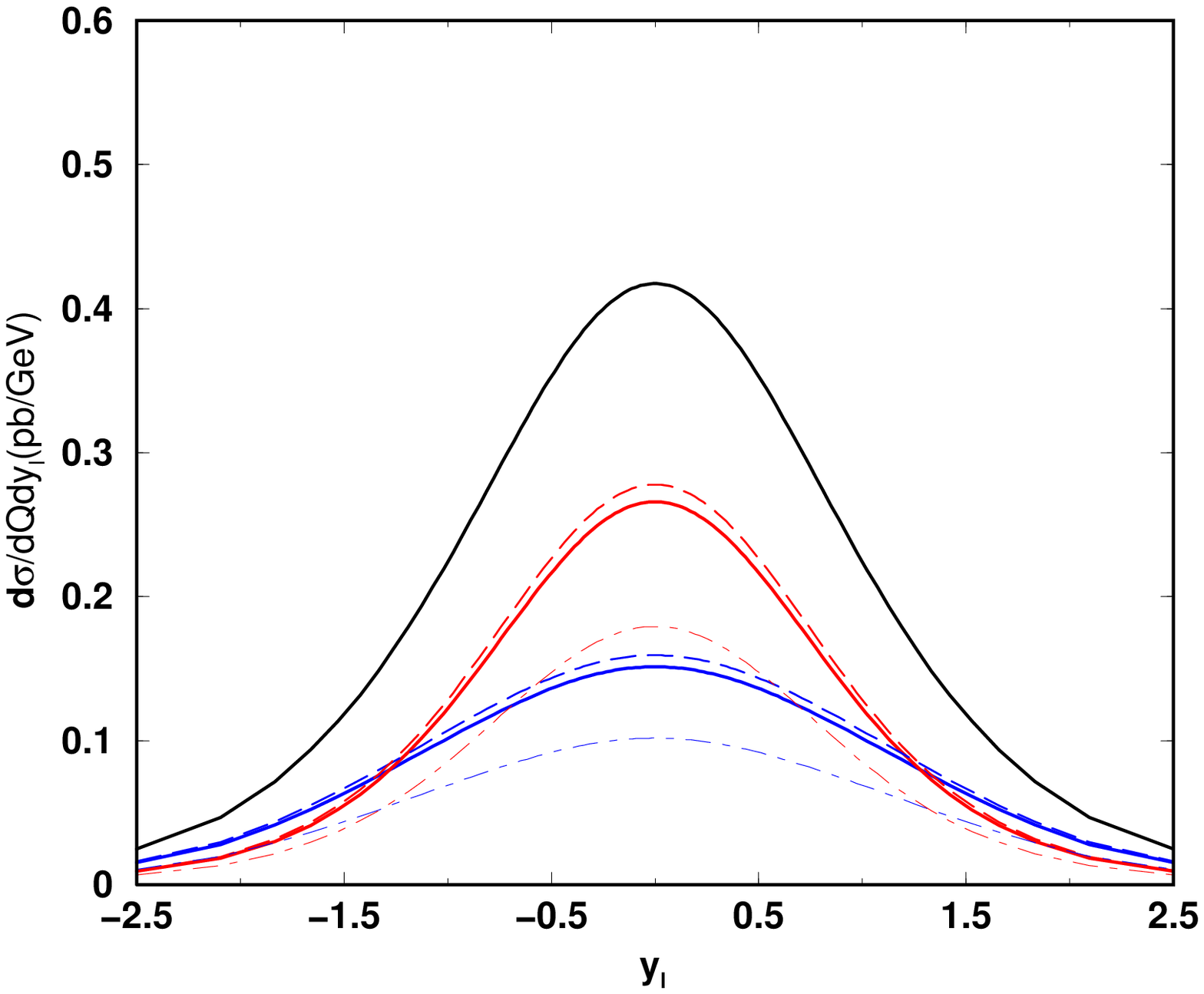,width=3.33cm} &
\leavevmode\psfig{file=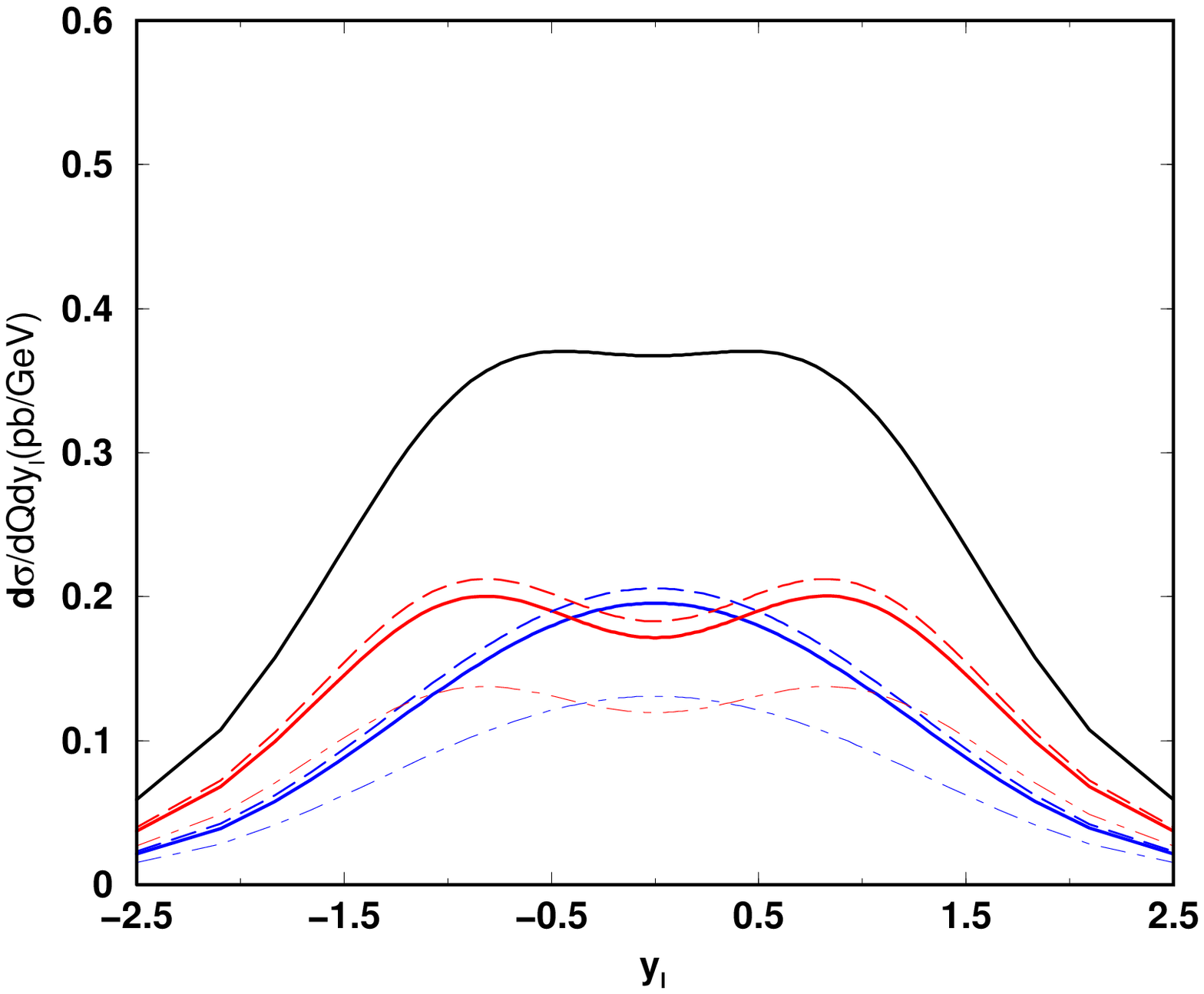,width=3.33cm}\\
{\footnotesize \ (3a)} &   {\footnotesize \ (3b)}
\end{tabular}
\vspace{-1cm}
\caption{Lepton distribution with $P_A (+) P_B (+)$}
\label{fig:3}
\end{center}
\end{figure}

\vspace{-1cm}
Figs. 3a (3b) and 4a (4b) show
lepton distributions with positive (negative) helicity for the
$P_A (+) P_B (+)$ and $P_A (-) P_B (-)$ configurations.
In the case of $P_A (+) P_B (+)$, $u \bar{u}$ and $d \bar{d}$
annihilations, roughly speaking, contribute similarly in size.
For $P_A (-) P_B (-)$ annihilation, $u \bar{u}$ sub-process
dominates the cross section.
One can understand these behaviors again intuitively
by noting the observation explained before concerning
Fig.1. 

\vspace{-0.8cm}
\begin{figure}[h]
\begin{center}
\begin{tabular}{cc}
\leavevmode\psfig{file=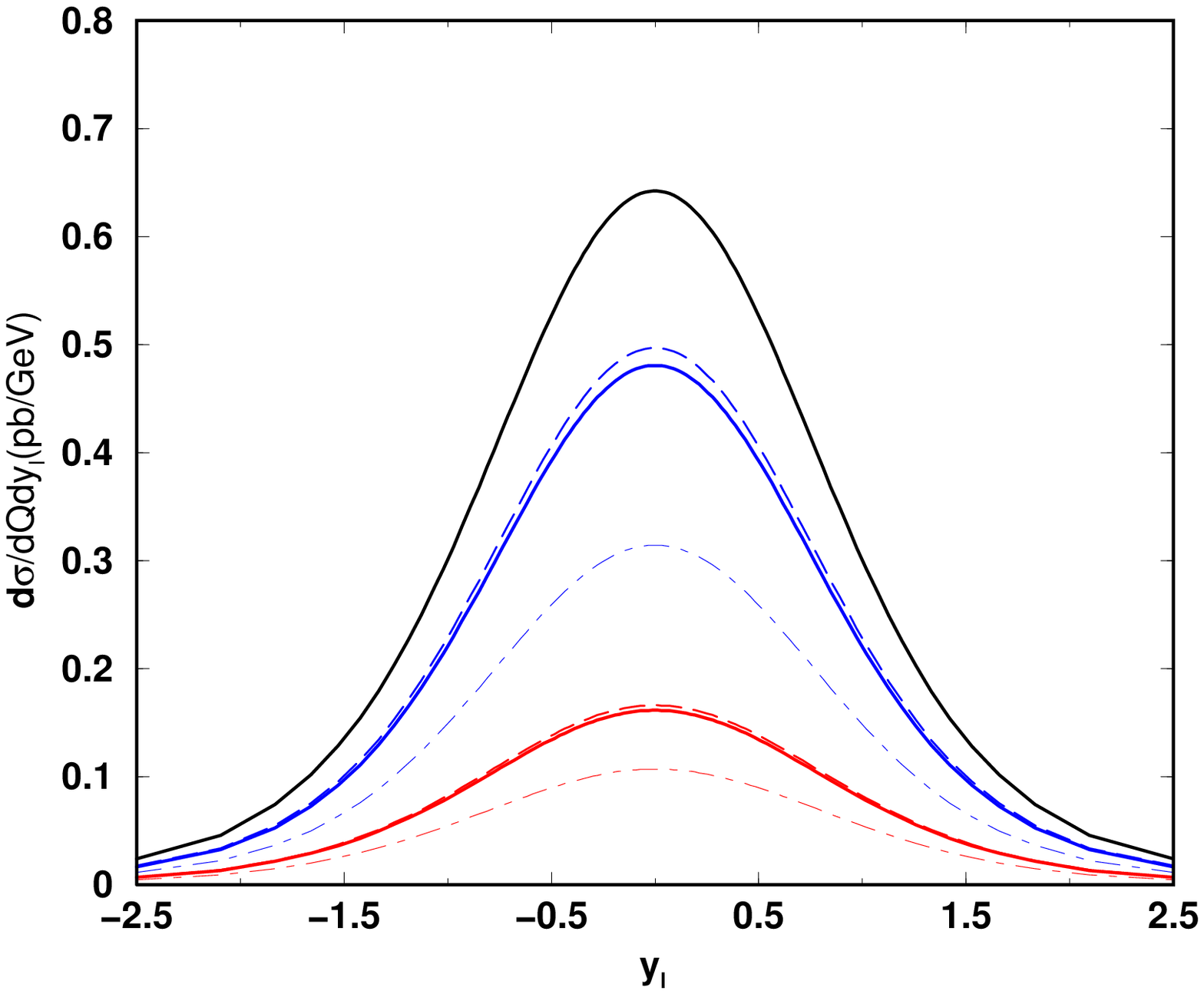,width=3.33cm} &
\leavevmode\psfig{file=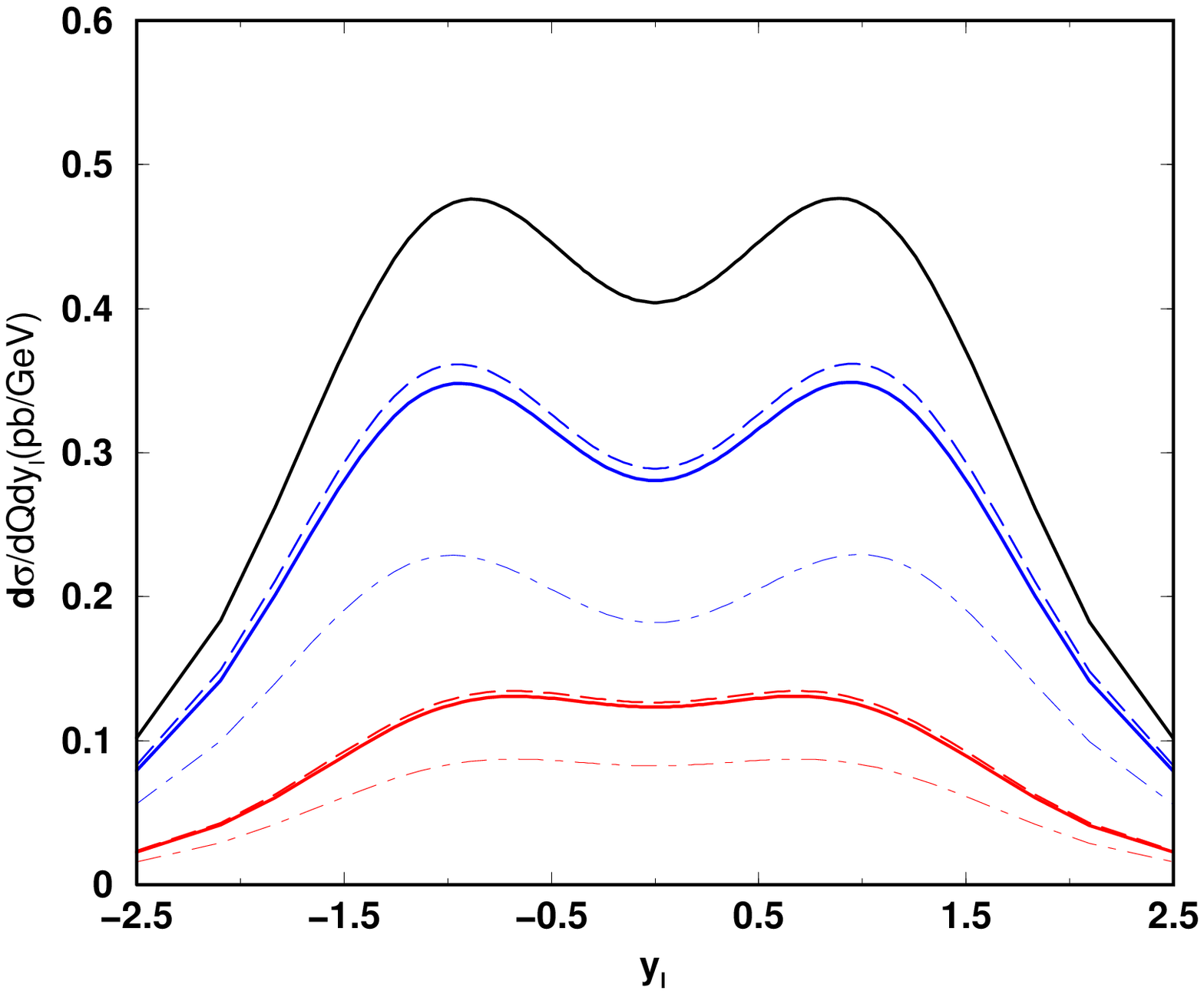,width=3.33cm}\\
{\footnotesize \ (4a)} &   {\footnotesize \ (4b)}
\end{tabular}
\vspace{-1cm}
\caption{Lepton distribution with $P_A (-) P_B (-)$}
\label{fig:4}
\end{center}
\end{figure}

\vspace{-0.8cm}
We also estimate the lepton helicity asymmetry $A$ which is defined by,
\[ A \equiv \frac{d\sigma(\lambda_l = -1) -d\sigma (\lambda_l = +1)}
    {d\sigma(\lambda_l = -1) + d\sigma(\lambda_l = +1)}\]
This asymmetry is plotted in Fig. 5 for the 
configuration of $P_A (+) P_B (-)$
using the various parton parameterizations in Refs.\cite{PDF1,PDF2}.

\vspace{-0.8cm}
\begin{figure}[h]
\begin{center}
\leavevmode\psfig{file=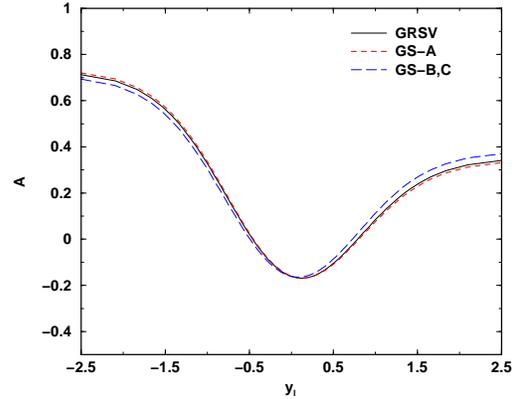,width=6.7cm}
\vspace{-1cm}
\caption{Helicity asymmetry distribution with $P_A (+) P_B (-)$}
\label{fig:5}
\end{center}
\end{figure}
\vspace{-1cm}
\noindent
This asymmetry amounts to around 50\%, but
the dependence on the various parton parameterizations is
quite weak.
The reason is that the process is dominated by the
quark and anti-quark annihilation in the RHIC energy region
and the gluon initiated Compton subprocess,
in which the ambiguity of gluon distribution will appear, 
gives a tiny correction to the cross section.
This is already expected from also the fact that
the QCD correction has mainly an enhancement effect of the tree
level cross section.

Figs.6 show the same asymmetry for the case of 
$P_A (+) P_B (+)$ (6a) and $P_A (-) P_B (-)$ (6b).

\vspace{-0.8cm}
\begin{figure}[h]
\begin{center}
\begin{tabular}{cc}
\leavevmode\psfig{file=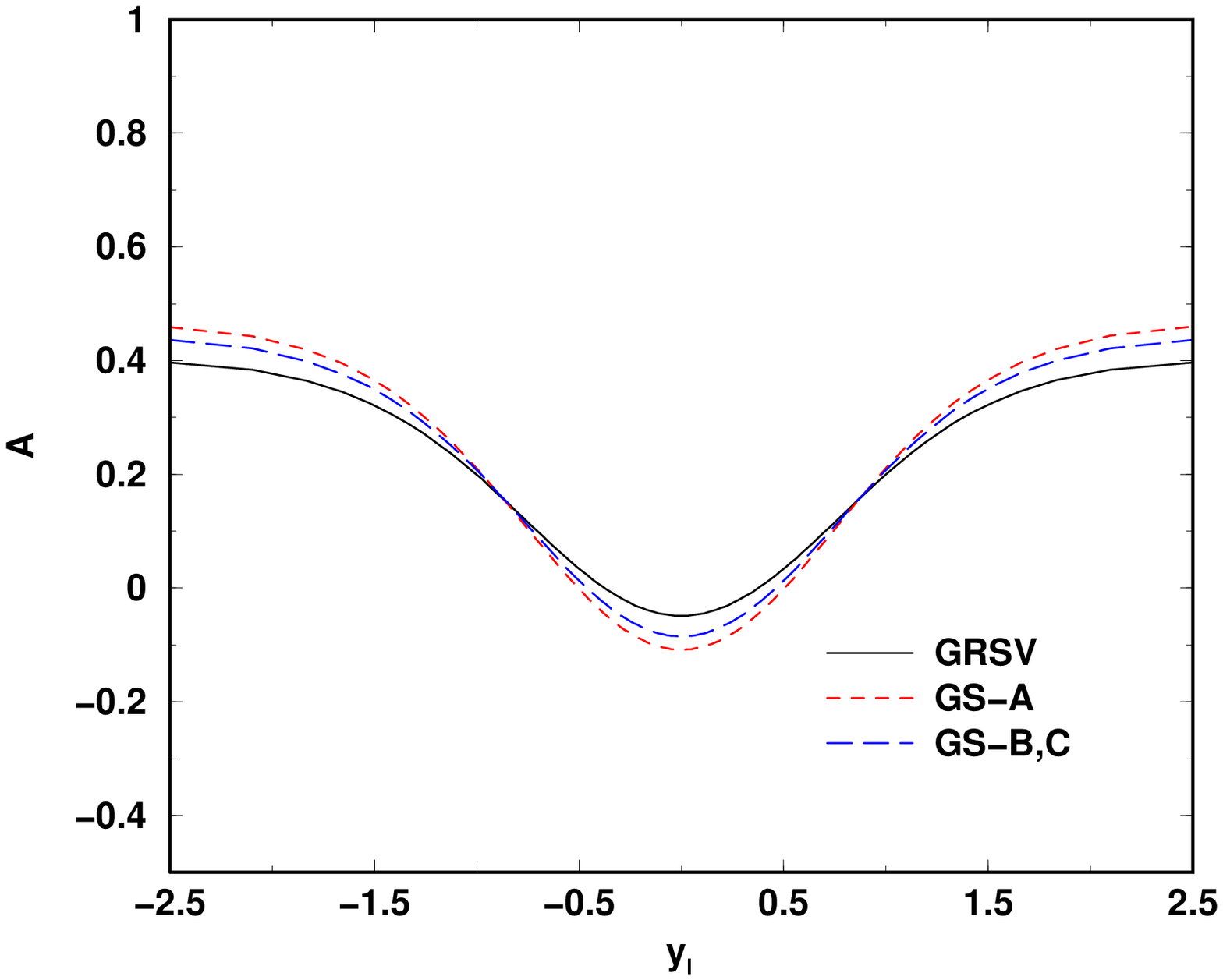,width=3.33cm} &
\leavevmode\psfig{file=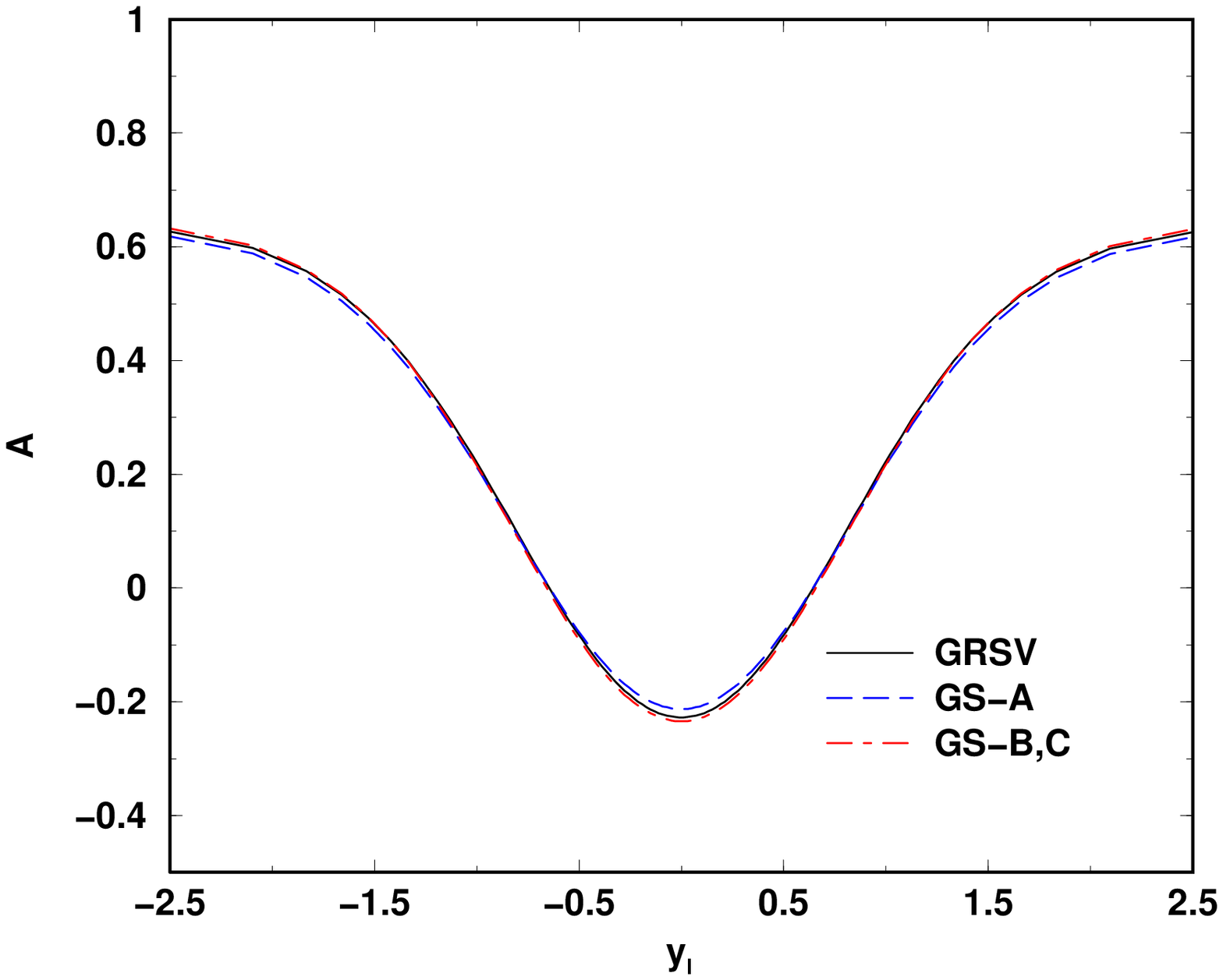,width=3.33cm}\\
{\footnotesize \ (6a)} &   {\footnotesize \ (6b)}
\end{tabular}
\vspace{-1cm}
\caption{Helicity asymmetry distribution with $P_A (\pm) P_B (\pm)$}
\label{fig:6}
\end{center}
\end{figure}

\vspace{-1cm}

\section{SUMMARY}

We have presented the lepton helicity distributions
in the polarized Drell-Yan process at the  ${\cal O} (\alpha_s)$
level in QCD.
We have numerically analyzed the cross section on the $Z$ pole
and pointed out that the $u (\bar{u})$ and $d (\bar{d})$ quarks 
give different and characteristic contributions to the lepton 
helicity distributions which deserve some theoretical
interests.
The QCD corrections mainly enhance the tree level
cross sections and this fact can explain qualitatively
the lepton helicity distributions from the various
proton's spin configurations. 

We have also estimated the lepton helicity
asymmetry which amounts to around 50\%.
Since the $q-\bar{q}$ subprocess is dominant in the RHIC
energy region, we will not be able, unfortunately, to
find difference between various parton parameterizations
which have big ambiguities in the gluon distributions.

From the experimental point of view,
it seems very difficult to measure the helicity of
produced muon and/or electron from Drell-Yan process.
However, if we can observe the $\tau$ lepton produced from
the Drell-Yan process and its decay,
we will be able to compare the experimental data and theoretical
prediction.

We hope that various kinds of new experiments
and theoretical investigations will be able to
clarify not only perturbative and nonperturbative
aspects of QCD but also the full structure of all interactions
in Nature.

The work of J. K. is supported
in part by the Monbu-kagaku-sho Grant-in-Aid
for Scientific Research No. C-13640289.


\end{document}